Musings on the theory that variation in cancer risk among tissues can be explained by the number of divisions of normal stem cells

## Cristian Tomasetti[1]* and Bert Vogelstein[2]*

"*He maketh His sun to rise on the evil and on the good, and sendeth rain on the just and on the unjust.*"
(The Bible, KJV - Matthew 5:45)

**Introduction** - Suppose an epidemiologist had discovered an industrial agent (Agent R) that was present in the normal tissues of every person in the U.S. at concentrations that were extremely highly correlated (0.80) with the risk of cancer in those tissues.  And suppose this agent was shown to be highly mutagenic.   S/he would have concluded that Agent R was likely responsible for ~2/3 of the variation in cancer risk among those tissues, more than all other environmental agents and heritable factors combined.  Given that cancer is known to be the result of mutations, the obvious interpretation of this discovery would be that a large fraction of the mutations responsible for cancer are caused by mutagen R. That scenario is conceptually identical to the actual one we reported in our paper.  But Agent R is not exogenous, it is simply normal cell division.  Every time a normal cell divides, ~3 mutations are created because of the imperfect nature of DNA copying. These mutations can be called "replicative" and are unavoidable, just as evolution is unavoidable. That they play a larger role in cancer than previously believed has important scientific and societal implications.

We recently published a paper (*Science* 347:78-81, 2015) suggesting that replicative mutations play a larger role in cancer than previously believed.  The data in this paper and their interpretation have important scientific and societal implications, and have thereby stimulated a healthy scientific debate.  The following points address several of the most important statistical and technical issues related to our analysis and conclusions.   Our responses to non-technical questions are available at http://www.hopkinsmedicine.org/news/media/releases/bad_luck_of_random_mutations_plays_predominant_role_in_cancer_study_shows

[1] Division of Biostatistics and Bioinformatics, Department of Oncology, Sidney Kimmel Cancer Center, Johns Hopkins University School of Medicine and Department of Biostatistics, Johns Hopkins Bloomberg School of Public Health, 550 N Broadway, Baltimore, MD 21205, USA.

[2] Ludwig Center for Cancer Genetics and Therapeutics & Howard Hughes Medical Institute, Johns Hopkins Kimmel Cancer Center, 1650 Orleans St, Baltimore, MD 21205, USA.

*Corresponding authors. E-mail: ctomasetti@jhu.edu; vogelbe@jhmi.edu

**Point 1 - Cancer subtypes and subgroups.** Cancers can be grouped by type (lung cancers vs. colon cancers), by subtype (e.g., adenocarcinomas vs. squamous cell cancers of the lung) or even further by subgroups within each subtype (e.g., lung adenocarcinomas in smokers vs. non-smokers). We included a total of 25 cancer subtypes and multiple subgroups in Fig. 1, for a total of 31 data points. For example, we included lung adenocarcinomas in smokers vs. non-smokers as two different subgroups, and liver cancers in HCV-infected individuals vs. non-HCV-infected individuals as another two subgroups. If we wished to consider only the variation in cancer risk among cancer subtypes in the overall population (i.e. no subgroups), it would have been preferable not to separate patients with the same subtypes into different subgroups. Indeed this separation was not included in the first step of our analysis (unpublished). In this case, by removing FAP and Lynch colorectal and FAP duodenum (since they are already accounted for in the lifetime risk of colorectal and duodenum adenocarcinoma, respectively), and grouping together head and neck squamous cell carcinomas (lifetime risk = 0.0145), hepatocellular carcinomas (lifetime risk = 0.00774) and lung adenocarcinomas (lifetime risk = 0.0276), the 31 data points reduce to 25 subtypes. The Spearman's rank correlation coefficient as well as the Pearson's linear correlation coefficient among these 25 cancer subtypes was 0.80 (0.59-0.91, 95% CI; $P < 1.8 \times 10^{-6}$). When we instead included the subgroups described in Fig. 1, Spearman's correlation coefficient was 0.81 and Pearson's was 0.80. We therefore felt comfortable including those subgroups in the final analysis, presented in Fig. 1. Their inclusion allowed us to illustrate the relative position of those subgroups in the graph and provided a segway to the subsequent discussion of ERS (Extra Risk Score).

In the points below, only the 25 cancer subtypes, ignoring subgroups, will be considered, due to the different questions addressed here, and as they are most germane to the overall relationship between stem cell divisions and cancer lifetime risk (hereafter referred to as cancer risk).

**Point 2 - Bad luck.** We define replicative mutations as those that occur as a result of errors made by polymerases during divisions of cells in the absence of any exogenous carcinogens. Such mutations occur at well-known rates that are very similar in all tissues that have been analyzed (Tomasetti et al., 2013). They are in part due to imperfect copying of DNA by DNA polymerases and incomplete correction of copying errors by DNA repair systems. When assessed at an overall level, they are randomly distributed among genes and individuals. Part of this randomness may be due to quantum effects during DNA replication (Lowdin, 1965) (Kimsey et al., 2015) or repair (Quesne et al., 2014). There are two other sources of mutations: those that are inherited (germ-line) and those that occur as a result of exposures to exogenous carcinogens. Most mutations, whether replicative, inherited, or carcinogen-induced, are innocuous, but a small fraction of them occur at specific positions within driver genes and thereby contribute to neoplasia (Vogelstein et al., 2013). We used the term "bad luck" to denote the effects of these replicative mutations for several reasons. First, replicative mutations are beyond our control. Replicative mutations are due to the same processes that drive evolution; copying of DNA is not perfect, even in a test tube under ideal conditions, and stem cell divisions are required for proper embryonic and post-natal development as well as tissue homeostasis throughout life. As Dr. Steve Elledge (Harvard) puts it, they are "the canvas, so to speak, upon which the picture is painted" (personal communication). Additionally, the term was chosen to distinguish replicative mutations from those caused by exposures to environmental factors, which are in principle preventable and not out of our control. The term "bad luck" could also be used to describe the effects of inherited mutations. However, inherited mutations are not included in our definition of "bad luck" because they are different from a mechanistic standpoint and because they could, in the future, be prevented through technologies now being explored.

We provide a new model for explaining cancer risk across tissues, introducing the number of stem cell divisions as a new fundamental variable. In our model, all other environmental and inherited factors contribute to cancer risk by increasing the probability of a mutation per stem cell division or by increasing the number of stem cell divisions - *but, by definition, environmental and inherited factors cannot increase replicative mutations*. Replicative mutations are the lower bound of mutational load expected from normal developmental processes, i.e., normal stem cell divisions. Our model provides the first mathematical quantification of the contribution of replicative errors to the neoplastic process. We interpret our results as indicating that replicative errors are responsible for nearly 2/3 of the variation in cancer risk across tissues (also see note about association and causation in Point 5). This estimate is derived from the square of the linear Pearson's correlation coefficient 0.80 (0.59-0.91, 95% CI; $P < 1.8 \times 10^{-6}$), and not from the ERS data in Fig. 2 (which showed a grouping of tumor types according to their ERS scores).

Could the correlation depicted in Fig. 1 of our paper be partly result of a connection between the number of stem cell divisions and the effects of environmental factors? We consider three categories of environmental factors: viruses, chemical carcinogens, and radiation. Viruses linked to cancer, such as HPV, HBV, HCV, HIV, EBV, HHV-8, are transmitted sexually or through other specific anatomic routes, such as needle-sharing. Thus, the cancers to which they contribute are determined by exposure and infectability of the corresponding normal tissues; these viruses do not infect tissues proportionally to their number of stem cell divisions. Overall these viruses do not tend to infect cells with high numbers of stem cell divisions.
Similarly, other carcinogens, such as smoking or dietary factors, affect tissues based on their level of exposure and many other factors but apparently not on their number of stem cell divisions. For example, there is no obvious relationship between the number of stem cell divisions and the relative excess risk of cancers of the head and neck, small intestine, esophagus and other exposed organs caused by alcohol (Bagnardi et al., 2001; Bagnardi et al., 2014).

Radiation provides a better answer to this question because it is less complicated by tissue-specific mutagen content. Mutagen "content" after some types of radiation varies only ~2-fold among most major tissues and can in fact be accurately quantified by radiation physics (Cullings et al., 2006). The most informative, albeit tragic, data on this subject have been accumulated through painstaking analysis of survivors of the Hiroshima and Nagasaki atomic blasts. Assume that environmental mutagens confer cancer risks to each tissue as a function of the total number of cell divisions in that tissue. Then a correlation between the number of stem cell divisions in tissue and the increase in risk for cancer in that tissue should be observed. Among the cancer types included in a study of cancer risk in atomic bomb survivors (Table 11 of (Preston et al., 2007), nine corresponded to those for which the lifetime number of stem cell divisions could be estimated. No correlation between the lifetime number of stem cell divisions and the excess absolute rate of cancer (EAR; Spearman's rho=0.13, P=0.74) or the excess relative cancer risk (ERR; Spearman's rho=-0.17, P=0.67) were observed (Figure TR1, a-b). In fact, for ERR the trend was negative. We also considered the possibility that a correlation might exist between the total number of stem cells in the tissue, rather than with the total number of stem cell divisions in that tissue, given the short time interval of the radiation exposure. However, even in this case, no correlation between the total number of stem cells and the excess absolute rate (EAR; Spearman's rho=0.03, P=0.94) or the excess relative risk (ERR; Spearman's rho=-0.18, P=0.64), was observed, and the trend was also negative (Figure TR1, c-d). Also the number of stem cells dividing during the time of irradiation, which is proportional to $s \cdot d$, see Supplementary Materials in (Tomasetti et al., 2015), does not correlate with the excess absolute rate (EAR; Spearman's rho=0.05, P=0.91) or the excess relative risk (ERR; Spearman's rho=-0.28, P=0.46). This analysis suggests that the correlation we found between cancer risk and total number of stem cell divisions is not due to the effects of environmental factors, but rather to replicative mutations.

The absence of a correlation of the lifetime number of stem cell divisions, the number of stem cells, or their division rate, with the tumorigenic effects of radiation does not mean that stem cells are not important for the effects of radiation. On the contrary, we offer the following analysis of these effects. Assume that the radiation caused by the atomic bomb, and administered within a time interval, increased the probability for one of the rate-limiting mutations required for cancer. For simplicity (and reality), let us assume that the radiation is responsible for the initiating mutation. If we disregard clonal expansions caused by fitness advantages, then the incidence of a given cancer type in the atomic bomb survivor group ($I_s$) would be approximately (for $ut \ll 1$)

$$I_s(t) = s\, u_1\, u_2\, \ldots\, u_n\, \frac{t^{n-1}}{(n-1)!}$$

where $s$ is the total number of stem cells in that tissue, $u_i$ is the mutation rate for the *ith* required driver mutation still needed to occur (n total), and $t$ is the time after the explosion, see ((Tomasetti et al., 2015) for more details). In a control group not exposed to the bomb, the incidence of that same cancer type would be

$$I_c(t) = s\, u_1\, u_2\, \ldots\, u_{n+1}\, \frac{t^n}{n!}.$$

Then,

$$\frac{I_s(t) - I_c(t)}{I_c(t)} = \frac{n - u_{n+1}\, t}{u_{n+1}\, t}$$

a decreasing power function of *t*, and

$$I_s(t) - I_c(t) = s\, u_1\, u_2\, \ldots\, u_n\, \frac{t^{n-1}}{(n-1)!}\left(1 - \frac{u_{n+1}\, t}{n}\right)$$

an increasing power function of *t*. This is indeed the overall behavior that is observed for ERR and EAR among atomic bomb survivors (Fig. 4, 8 of (Preston et al., 2007)). Thus, these data are consistent with the idea that radiation emitted by the atomic bomb induces cancer by increasing the probability of a rate-limiting mutation in a stem cell. This can explain the timing of the extra cancers in individual tissues but not the variation in cancer risk among those tissues (Fig. TR1). Evidently, tissue-specific factors are involved in mediating the mutagenic effects of this sort of radiation (high doses, short exposures). Similar tissue specific factors are likely to mediate the effects of other environmental agents, such as alcohol in the example noted above. In contrast, the replicative mutations that occur during normal stem cell divisions appear to be "mediated" similarly among all tissues, or the correlation observed in Fig. 1 would not have been observed.

It is also worth pointing out that our correlations, though high (~0.8), were not perfect. The slope of the regression line was also not 1.0. This means that there is not a 1:1 relationship between stem cell divisions and cancer risk (in log scale). Many factors could account for these variations. While the number of stem cells present in a tissue should scale linearly with respect to cancer risk in that tissue (double amount of cells should yield a double risk), cancer risk should scale log-linearly with the total number of divisions of each stem cell lineage during a lifetime. The specific slope is also a function of the number of required driver events needed for cancer to occur. Because we do not know the number of required drivers for each tissue analyzed, our ability to correct for it is limited. For example, some cancers appear to require only one driver mutation, while others require three (Tomasetti et al., 2015; Vogelstein et al., 2013). The inclusion of heterogeneity in the number of required drivers among different cancer types easily yields a slope equal to 1 (see Figure TR2). Our theory could also not explain certain important epidemiologic observations, such as why glioblastomas are more common than small intestinal cancers; the number of normal small intestinal stem cell divisions is much greater than those of normal glial stem cells. This indicates that our estimate of the number of stem cells related to glioblastomas or small intestinal cancers is incorrect, that there are other factors at play (such as a powerful immune protective response in small intestine), or that our theory is flawed. Finally, our estimation for the explained variation depends on the tissues we were able to include in our analysis. The estimate may change as the dynamics and characterization of stem cells in more organs becomes available, thereby allowing the inclusion of cancer types in these organs in our model. However, the strong correlation we found is very robust (see Material and Methods), and **an important component of variation in cancer risk appears to be explicable by the number of stem cell divisions.** Figure 1 presents a striking relationship, extending over more than five orders of magnitude, never before observed between cancer incidence and any other factor.

**Point 3 - Log-log transformation.** Our analysis was designed to determine the proportion of the variation in cancer risk (aka, lifetime cancer risk) among tissues that could be explained by the total number of stem cell divisions. The distribution of these cancer risks is extremely skewed in the original scale, see Fig. 1 in (Tomasetti and Vogelstein, 2015). Thus, the standard regression applied to the data in the original scale would not be appropriate. The log-log is the appropriate scientific scale to use for such analysis. We therefore used a $\log_{10}$ transformation to estimate Pearson's correlation coefficient and $r^2$. This is standard practice in statistics.

Suppose we had not transformed the data to log-log scale? Pearson's correlation coefficient would then be much higher, rather than lower. In fact Pearson's correlation coefficient would be 0.96 (0.92-0.99, 95% CI; $P < 5.2 \times 10^{-15}$), with $r^2 = 0.93$. This higher coefficient results from the overwhelming contribution of the few cancer types with the highest cancer risk. We therefore purposely chose the more conservative estimate, based on a log-log transformation of the data.

As noted above, we conclude that about 2/3 (0.35-0.83; 95% CI) of the variation in the lifetime cancer risk ($\log_{10}$) across tissue types in the U.S. could be explained by the lifetime number of stem cell divisions ($\log_{10}$). Note that with or without the log transformation, Spearman's rho is 0.80, implying that about 2/3 of the variation in the *rank* of the various cancer risks could be explained by the *rank* in the number of stem cell divisions in the stem cells giving rise to each cancer type. These data indicate that there is a fundamental, large association between these two variables (cancer risk and number of normal stem cell divisions), whether untransformed or log-transformed data is used to measure it.

**Point 4 - Absolute vs. relative risk.** In some media reports, the difference between *absolute* and *relative* cancer risk was not fully appreciated. As noted above, our results relate only to relative variation in cancer risks, not to absolute variation in cancer risks. In other words, we sought to understand *why* hepatocellular cancers occur 26-fold more commonly than duodenal cancers rather than to understand why duodenal cancers occur in ~1 in 3,333 individuals in the U.S.

Can we estimate the contribution of stem cell divisions to the *absolute* risk of each cancer type? Though not included in our paper for reasons discussed below, it is possible to attempt doing this from the data in Fig 1. As explained in Point 3, we used a log-log transformation of the data. With data on a log scale, there is obviously no zero on the y-axis. However, if we extrapolate from Figure 1, the linear regression yields an intercept of approximately $10^{-7}$. If one were willing to accept that this baseline value ($10^{-7}$) is essentially zero, the extrapolation would imply that 64% (i.e., the square of Pearson's correlation coefficient) of the overall variation (*absolute*) in cancer risk ($\log_{10}$) of developing cancer of the types examined could be explained by the lifetime number ($\log_{10}$) of stem cell divisions. This extrapolation seems reasonable because the baseline value would correspond to only about 30 cancer cases among the entire U.S. population[1].

Another way to estimate the contribution of stem cell divisions to the *absolute* risk of cancer, and one that avoids the dependence of the relationship on log10 values, can be made through the use of weighted regression. As we wished to explain the variation in cancer risk across tissues in our paper, it would have been inappropriate to give more weight to one cancer type than to another. But such weighting on a log-log scale can indeed be used to provide an estimate of the absolute risk in the actual, untransformed scale (Delta method). In brief, it is possible to perform a weighted regression of the log-transformed data with the weights given by the absolute cancer risks of each cancer type. By using the log-log scale for this regression, we avoid the skewed distribution issues noted above. Such an analysis reveals that 68.4% of the *absolute* risk (in the actual scale) of developing cancer of the types examined can be explained by the lifetime number of stem cell divisions. This means that ~2/3 of the actual cancer incidence (as explained by the squared variation) for the cancers types included in the analysis may be explained by the lifetime number of stem cell divisions, and therefore by replicative mutations. This conclusion assumes that environmental and inherited factors are uniformly distributed in the population, and incorporates the evidence relating stem cell divisions to replicative mutations reviewed in Point 2.

One of the reasons we did not include this estimate of the proportion of *absolute* risk assignable to stem cell divisions in our paper is that only 25 tumor types were analyzed. There are many other tumor types, including very common ones such as prostate and breast cancers, which we could not include because of insufficient data on their normal stem cells' identity or dynamics. However, according to Cancer Research UK, no prostate cancer cases and only 21% of breast cancers are thought to be preventable.

**Point 5 - Bad luck and prevention:** We are not the first to estimate the proportion of cancer risk

---

[1] During the preparation of this report, Dr. Bernd Fritzsch (Professor of Biology and Director of Center on Aging, University of Iowa), informed us that the inner ear provides an example of a tissue that must have extremely few stem cells. He estimates that there are many fewer than 60,000 stem cells because there are only ~15,500 hair cells and 35,000-50,000 spiral ganglion neurons (all post mitotic) in the inner ear. According to Dr. Fritzsch, no cancers of the sensory epithelial cells of the cochlea have ever been reported.

attributable to "bad luck", though the issue is not usually framed in this way.  The proportion of cancer cases due to bad luck (aka "unpreventable" or "beyond our control") has been estimated by the U.S. CDC as ~80% and by Cancer Research UK as ~60%.  The data in our paper do not allow us to estimate the proportion of the total cancer cases that are preventable, because that would require knowledge about all cancer types and calculations of absolute risks subject to the caveats above.  On the other hand, our data are intuitively consistent with the data from either the CDC or Cancer Research UK that 60% to 80% of cancers in such developed countries (and, to be more specific, in the U.S.) are "unpreventable". More importantly, our data and interpretation provide a mechanism - replicative errors - that helps explain why many cancers are "unpreventable"

Though our estimates of absolute risk are thus in excellent agreement with these others, there are fundamental differences about the approaches used and the interpretation of these estimates.   Most importantly, the previous estimates are based on comparisons of cancer cases among different groups of individuals in a wide range of environments and with varying genetic constitutions.   Though certain diets are known to be associated with lower or higher risks, the actual dietary components are often unknown.   In contrast, our estimates are based on two parameters - the number of stem cell divisions and the cancer risk of different tissues. Given these differences, the agreement between these estimates is remarkable.
The mechanisms responsible for the cancer types that did not appear to be preventable was heretofore obscure.   They could be based on as yet undiscovered exogenous agents, such as carcinogens or tumor viruses.  They also could be based on a variety of stochastic process, such as those involved in determining whether a given tumor elicits an immune response in the individual, or whether some epigenetic switch is thrown (Segal et al., 2008).   Our results point to a particular stochastic process (replicative mutations) occurring during normal cell divisions as a major culprit.  We are of course aware that our theory is entirely based on a correlation, and correlations cannot be equated with causation.  However, we believe that our theory is strongly supported by the voluminous data indicating that somatic mutations are responsible for cancers in general (Vogelstein et al., 2013) and by equally compelling data showing that normal cells make random errors at specific, uniform rates when they replicate their DNA.

As with any correlation, ours can be compromised by unappreciated confounders.  For example, it is possible that the mutation rates in every normal stem cell are largely due to some unappreciated carcinogen that is present at roughly equal concentrations in every normal stem cell type in the body. We feel this confounder is possible but implausible given that all known carcinogens are cell type specific.  For example, ultraviolet light is a major carcinogen but its "concentration" is much higher in the skin than in the gallbladder.   Moreover, the unappreciated carcinogen would have to be avoidable if it were to play a role in prevention.   An example of this is oxygen, to which all tissues are exposed, at roughly equal concentrations.  Oxygen, through the production of reactive oxygen species, contributes to mutation rates.  But exposure to oxygen is largely unavoidable and we consider that any of its effects are already incorporated into the known rates of mutation in normal cells.

Any discussion of preventability should include primary prevention and secondary prevention.  Primary prevention is implemented by measures that reduce the incidence of cancer, so that cancer is never diagnosed. These measures include changes in lifestyle and vaccination against cancer-associated viruses.  Secondary prevention is defined as measures that decrease cancer deaths once cancer has occurred, such as can sometimes be accomplished through early detection of disease. Though the incidence of many cancers, and many cancer types, might not be preventable through primary prevention, secondary prevention has the capacity to reduce deaths from nearly all forms of cancers. However, our analysis does not diminish in any way the importance of primary prevention on reducing cancer risk. Primary prevention is always preferable to secondary prevention when the former is possible.

**Point 6 – ERS and clustering:** From examination of Fig. 1, it is obvious that some cancers are outliers from the regression. We suspected that the more a cancer type deviated from the regression line, the more likely that other factors, such as environmental exposures or heritable genetic variations, played a role in its pathogenesis. To evaluate this suspicion, we developed the ERS as a first attempt to measure these deviations. The 22 cancer types and nine subtypes were then clustered into two groups in an unsupervised fashion based only on their ERS. As shown in the figure below, this clustering showed that one of the two groups was largely composed of the tumor types and subtypes that were well known to be associated with specific environmental or hereditary influences. This group included lung cancers in smokers, hepatocellular cancers in HCV-infected individuals, and colorectal cancers in patients with mismatch-repair deficiency. We considered it surprising that an extremely basic biologic feature, unrelated in any obvious way to cancer, could not only predict which cancer types would be more prevalent but also which cancer types would more likely be the result of exogenous rather than endogenous factors.

We are aware of the fact that the ERS values do not show a bimodal distribution, and therefore do not cluster into two natural groups. If we had clustered into more than two groups, we would have obtained more than two clusters, with various degrees of evidence for extra risk factors, from low to high. A benefit of using two clusters was that it allowed a straightforward classification of the tumor types and subtypes into D-tumors (D for deterministic) and R-tumors (R for replicative). We suggested that environment and hereditable genetic variation played a greater role in D-tumors than in R-tumors. We note here, however, that a cancer type classified as an R-tumor does not have to be only, or predominantly, the result of replicative mutations. It is simply a cancer type for which a classification into two clusters placed it in the cluster where there is less evidence for the role of cell division in its pathogenesis. The ERS provides a value for each cancer types based on the evidence for extra risks, i.e., the effects of environmental or hereditary factors, to assess the evidence for extra risk factors (environmental and inherited) in a relative way, by comparing the value assigned to each cancer type across tissues.

The ERS provides a continuum of values, and the position of a given cancer type or subtype with respect to other cancer types may prove informative, particularly as better estimates of stem cell dynamics become available. Additionally, ERS is a measure of the evidence for extra risk in a cancer type in the overall population and does not convey any information about the possible existence of subgroups of individuals in which that cancer type is strongly affected by environmental or inherited factors. There may be a relatively small group of patients with a particular cancer type wherein environmental or inherited factors play a large role in that cancer's risk. Such subgroups would not affect the ERS for the overall type because the scale used for the analysis across cancer types is different from the scale that would be used to analyze subgroups within a cancer type.

Some further comments about the rationale behind the ERS are provided here. Consider Figure TR3 below. As in Fig. 1 of the paper, we assume that a given cancer tissue type is represented by a point in that figure, such as the grey and green points. In one case (green point below the grey point), the lifetime risk is much smaller even though the number of stem cell divisions is the same. In the other case (green point to the right of the grey point), the number of stem cell divisions is much larger but the lifetime risk is the same. The key idea behind the ERS is that both cancers represented by green points have less evidence for the effects of environmental or inherited factors than the cancer represented by the grey point. This is reflected by the fact that the areas of the rectangles formed by the green points with the coordinate axes are larger than the area of the rectangle formed by the grey point (Fig.TR3). Similarly, the cancer type represented by the blue point provides greater evidence for the effects of environmental or inherited factors than the cancer type represented by the grey point, since the cancer type represented by the blue point has a much higher risk but the same number if stem cell divisions. And indeed the area for the blue rectangle is much smaller than the area of the corresponding grey point. The product of the two coordinate values (in log10) for each cancer type equals the area formed by that point with the coordinate axes. And we use this area as a score for the evidence of extra effects on cancer risk (relative to other cancer types), beyond replicative mutations. The smaller the area, the higher the ERS score, as the product of the coordinate is always negative (logrisk is always <0 and stem cell divisions are always >0). It is important to note that each point can be considered to divide the entire 2-dimensional space represented in the log-log plot into 4 separate regions (red lines intersecting at the green point, Fig. TR3). All points contained in the top left region will "dominate" that green point, in the sense that the evidence for extra risk of the points in the top left region is larger. All points in the lower right region are instead "dominated" by that green point, in the sense that the evidence for extra risk in the bottom right region is smaller. All of the above is true assuming homogeneity across tissue types in the number of driver mutations required for cancer. Heterogeneity in the number of required driver genes confounds this analysis, which is why ERS and RBERS represent only a first attempt to assess the the relative evidence for environmental and inherited factors affecting a tissue type.

For the top right and bottom left regions, the relationship is less clear, and it will depend on the exact slope of the regression line derived from regressing on all the cancer types analyzed. As a first step, we wished to have a score that did not use any information derived from the regression because the regression is dependent on the number and types of cancer tissues included in the analysis. We therefore used the ERS as defined in the paper. However, based on the regression, the ERS score can be adjusted to obtain a new score, the RBERS (Regression Based ERS), that allows the assignment of the same score to points lying on a line with the same slope as the regression line (dotted line). Intuitively, the larger the vertical distance of a given data point above the regression line (its residual), the more likely that environmental or hereditary factors play a role in that cancer type. Thus, this distance can be used as a measure of the influence of those factors. We call the regression line the "replicative risk function" (RRF) since it may be used as a measure for the risk attributable to replicative mutations. Letting *RRF(lscd)* be the y-coordinate log10 value that the RRF takes in the log-log regression when the lifetime number of stem cells divisions is given by *lscd*, then we define, for a tissue with lifetime risk r and lifetime number of stem cells divisions *lscd*,

$$RBERS\ (lscd,\ r)\ =\ \log_{10} r - RFF(lscd).$$

The RBERS has advantages over the ERS. However, the RBERS has a major disadvantage: it imposes a straight-line relationship as the true relationship. This may not be correct, and its validity also depends on a slope that may change as more points are added to the regression line in the future. In contrast, the ERS is "unsupervised" - it uses no information from the regression analysis of the dataset and only employs the raw data. For each point, the ERS is completely independent of the values assigned to other points. Overall, both ERS and RBERS provide useful information and, importantly, they yield comparable results.

**Point 7 – Individual variation and multifactorial disease.** We have not considered individual variation within each cancer type. Given different lifestyles, environments and inherited genes, the risk of cancer will vary among individuals. Every individual cancer is the result of a possible combination of bad luck, "bad" genes, and "bad" environment, i.e., replicative mutations, inherited mutations, and mutations induced by exposure to carcinogens. Without knowing the details of a particular cancer, it is impossible to know the relative contribution of each of these three factors to a specific cancer. Referring to an analogy we have made elsewhere (see our Addendum to the JHU news release), consider an auto accident over a long trip in a car with a drunk driver, bad tires, and poor brakes who was driving in torrential rain and hit by an out-of-control tractor-trailer. This accident was not "caused" by any single factor but to some combination of these factors.

**Point 8 - Integrating our theory with molecular genetics.** It may be useful to extend our theory to include information about the molecular genetics of cancer. Most common cancers require a series of sequential mutations in driver genes to develop. We have shown in (Tomasetti et al., 2015) that the number of driver mutations will be at most 3. The small slope found in Figure 1 (~0.44) of (Tomasetti and Vogelstein, 2015) provides further evidence for this claim, i.e. that only a very few driver hits are required, since the higher the number of required drivers the higher the slope would have to be. In fact, in some cancers with lower $lscd$, even fewer driver mutations are sufficient, like in chronic myeloid leukemia, where only one chromosomal translocation is needed. The lack of correlation between the effects of environmental factors on cancer risk and $lscd$ (see point 2), as probably many other biological mechanisms (e.g. immune system), and even tissue organization and geometry, affect this slope. For simplicity, let us assume that the number of required driver hits is 3. Our extended analysis suggests that for those 2/3 of cancer cases that are due to replicative mutations, all three mutations are replicative. These mutations can occur in a normal stem cell or in an initiated cell, but occur at the rate expected for normal cells in the absence of exogenous influences. For the other 1/3 of cases, at least one mutation of the three is a non-replicative mutation. Thus, at least two of the three mutations, on average over all cancer types studied, may arise as a result of replicative mutations and the rest will result from mutations resulting from hereditary or environmental factors. Moreover, the ERS component of our theory suggests that the relative proportion of replicative mutations varies by cancer type. In cancer types with high ERS (Extra Risk Score; Fig. 2 of Tomasetti), such as lung cancers in smokers, two of the three mutations may be due to cigarette smoke and one due to normal replication. In other cancers, such as lung cancers in non-smokers or bone cancers, all three mutations are more likely to be replicative in nature.

**Point 9 – Response to Noble et al. manuscript (as posted on arXiv.org).** We are delighted that so many scientists from different disciplines are considering the issues brought up by our study. Below, we share our opinion on some of the major points raised in the Noble et al. manuscript, as posted on arXiv.org.
We appreciate the attempt to "extend" our analysis by separating the total number of stem cell divisions occurring in a lifetime of a tissue (**lscd**) in two separate components: the number of stem cells in a tissue (**s**) and the number of tissue self-renewals in a lifetime plus 1 (**d+1**). The authors indicate that this separation may further clarify the role of those two components: tissues with large s and small d may be potentially more affected by environmental carcinogens than tissues with the same lscd, but with small s and large d, where cancer may result mainly from the random replicative component. However, we believe that there is a fundamental biological problem with separating the total number of stem cell divisions in this way and attributing a larger part of the effects on cancer incidence to s, the size of the stem cell population in that tissue, as the authors do. It is impossible to separate these two components for the following reasons:

1) For many tissues (e.g. brain, lung) a large fraction of the divisions occur during the development phase of the tissue rather than during adult homeostasis. Thus, in those cases, it is s (specifically 2s) that accounts for most of the total number of stem cell divisions (and therefore for the random replicative mutations) rather than d. Thus in these cases, the conclusions would be the opposite than those indicated by the authors.

2) The total number of divisions during the self-renewal phase (i.e. after tissue development) is given by d·s. Some tissues have large d and small s, others small d and large s. What matters with respect to this phase is the product d·s, not just d. Only their product tells us how many divisions occurred and therefore how many opportunities for random replicative mutations occurred.

We have further addressed in our point 2 of this Technical Report the relationship between environmental effects and stem cell divisions. In brief, we provide more evidence that the environment does not appear to contribute to the total number of cell divisions in a way that is compatible with the data reported in Fig.1 of our paper.

We find also serious mistakes with the idea that the slope within each group of tissue types should be 1. It is true that one would expect a tissue double in size to have double risk (slope =1), but only if all other things were kept the same, which is not the situation here. For example, the frequency of divisions among different tissues of the GI tract, or between different pancreatic cells, is not constant. Thus, the measure of cancer risk per stem cell division proposed is flawed because it is based upon a slope equal to 1. We are currently considering other factors that may explain the slope of the regression and will describe this work elsewhere.

**Point 10 – Response to Altenberg's manuscript (as posted on arXiv.org).** Below, we share our opinion of some of the major points raised in the manuscript by Altenberg.

The ERS is the very first measure to assess the evidence for extra risk factors (environmental and inherited) in cancer incidence where the effects of random replications are taken into account. The criticism that ERS is not "invariant under a change of time units" is incorrect. Altenberg mistakenly refers to the fact that, in general, given some positive $T \neq 1$:

$$ERS(r, lscd) = \log_{10}(r) \cdot \log_{10}(lscd) \neq \log_{10}\left(\frac{r}{T}\right) \cdot \log_{10}\left(\frac{lscd}{T}\right) = ERS\left(\frac{r}{T}, \frac{lscd}{T}\right).$$

But this has nothing to do with time invariance and it is true for any equation when units are not taken into account. For example, the formula for the average speed of an object provides different values depending on the unit: the same speed, 360 km/hour, yields a value equal to 360 (km/hr) or equal to 6 (km/min), depending on whether time is measured in hours or minutes. Specifically,

$$s(d, t) = \frac{d}{t} \neq \frac{d}{60 \cdot t} = s(d, 60 \cdot t)$$

where *s* is the average speed of an object that travelled a distance *d* in kilometers in the interval of time *t*,

with time measured in hours.

In order to test for time invariance, it is not enough to simply divide by some constant T, but it is necessary to keep into account the time units. The ERS formula has a time unit (lifespan) that is implicitly included in the formula. Thus, to properly check for time invariance it is necessary to keep this fact into account. Without properly adjusting for the time units, even the measure that Altenberg considers appropriate and he ends up using, i.e. our other proposed measure RBERS (called by him RR, considered by us long before our Science publication, submitted before his comment, and included in the first version of this Technical Report, see Point 6), is not invariant under this type of transformation, since

$$log_{10}(r) - a - b \cdot log_{10}(lscd) \neq log_{10}\left(\frac{r}{T}\right) - a - b \cdot log_{10}\left(\frac{lscd}{T}\right)$$

for b≠ 1, where *a* and *b* are the coefficients of the regression line, i.e. the "replicative risk function" (RRF, see Point 6) obtained with respect to the original time units (/lifespan) in the original plot. If instead the time units are properly adjusted, a new regression line is obtained in a new plot with new time units. Geometrically, by changing the time units, both the data points and the regression line are shifted accordingly, in a new plot with new time units, keeping RBERS invariant. In the same way, when the time units are properly adjusted, both the data points and the ERS contour lines are shifted accordingly, in a new plot with new time units, keeping ERS invariant.

Mathematically, if lifespan equals 80 years, then

$$ERS_{year} = log_{10}(r_{year}) \cdot log_{10}(lscd_{year}) =$$

$$= log_{10}\left(r_{year} \cdot \frac{80\ years}{1\ lifespan}\right) \cdot log_{10}\left(lscd_{year} \cdot \frac{80\ years}{1\ lifespan}\right) =$$

$$= log_{10}(r_{lifespan}) \cdot log_{10}(lscd_{lifespan}) = ERS_{lifespan}$$

where the next to last equality is true because

$$r_{year} = r_{lifespan} \cdot \frac{1\ lifespan}{80\ years}, \quad lscd_{year} = lscd_{lifespan} \cdot \frac{1\ lifespan}{80\ years}$$

Thus, it follows that ERS is both time invariant and also monotonic, or rank-preserving, under this type of transformation. Thus, this criticism is completely invalid.

Also, not knowing the truth, that is, not knowing the shape of the true contour lines, there is no *a priori* reason for preferring the RBERS versus the ERS. Moreover, it is not "fortuitous" that ERS and RBERS provide somewhat comparable results for the relevant range of r and lscd in the log-log plot. It is because the ERS contour lines are close to linear in that region, something we were well aware of. Finally, Altenberg's "fallacy" section is based on a misunderstanding due to a typo, corrected by us before publication, but not included in the initial version of the publication. The typo has been corrected by Science, as noted below.

**Point 11 - Typos:** Some of our final edits of the proofs did not make it into the first printed version of the paper. Specifically:
- for glioblastoma, the label and value (-3.93) overlapped in Figure 2;
- the last sentence of the report should read: "For R-tumors, primary prevention measures are not likely to be *as* effective, and secondary prevention should be the major focus";
- in the Supplementary Material "incidence" should be "lifetime risk";
- in the Overview section of the Supplementary Materials should read "We assessed a total of 31 different cancer types. Other cancer types could not be assessed, largely because details about the normal stem cells maintaining the tissue in homeostasis, or their division rate, are missing";
- in the Overview section of the Supplementary Materials the following sentence should be added: "The uncertainty in the identification of the stem cell population and in the estimates for the number of stem cells, and their division rate, varies across the tissues we considered";
- the next to last section of the Supplementary Materials is should read: "For example, if the number of clusters was set to three, the cluster defined as "**R**-tumors" splits in two, one of which contains cancer types where environmental or inherited factors are known to play a role, and another where these effects are not known to play a role";
- Supplementary Figure S1 (a newer version, attached at the end this report).

*We thank Tom Louis (Professor of Biostatistics, Johns Hopkins Bloomberg School of Public Health and Chief Scientist, U.S. Census Bureau) for helpful discussion and comments.*

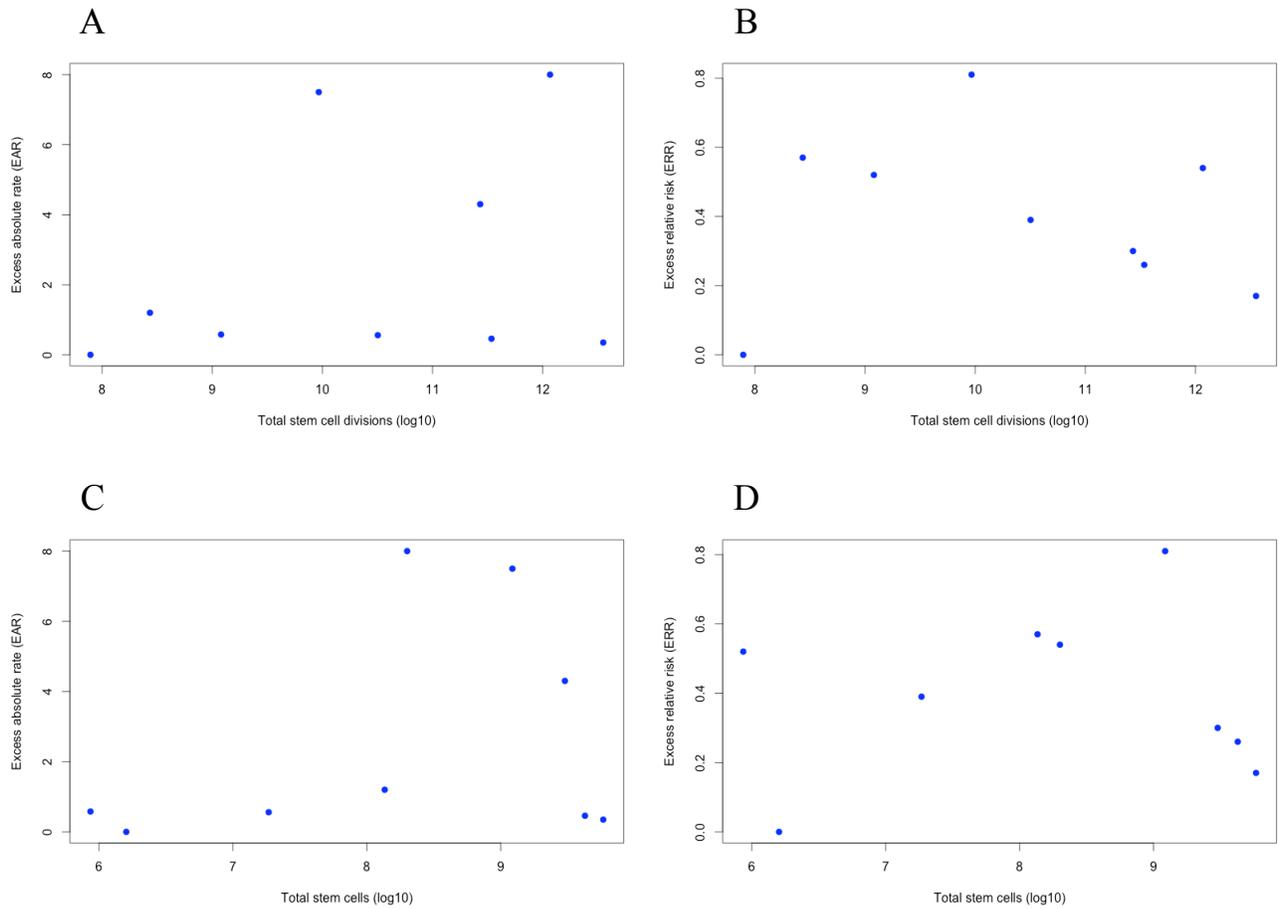

**Figure TR1. The effects of the atomic bomb radiation on tissues.** No correlation was found between the excess absolute rate (EAR) or the excess relative risk (ERS) caused by the radiation and the number of stem cell divisions (A-B) or the total number of stem cells (C-D). The 30-year old cohort is the one used in the analysis since more data is available on that cohort than any other.

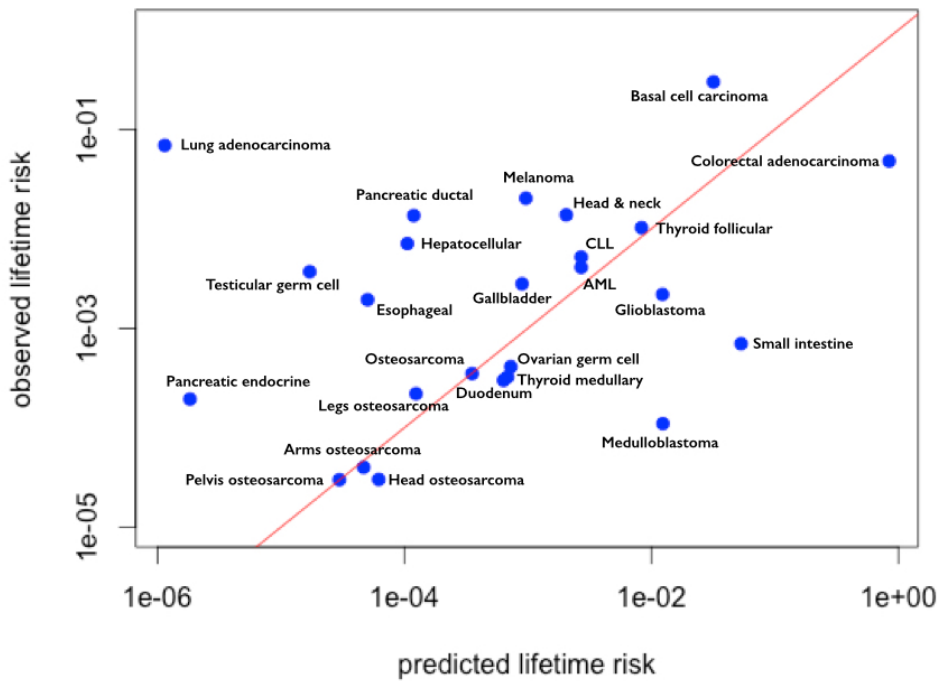

**Figure TR2. Theoretical predictions of cancer risk based on replicative mutations only versus the actual observed risks.** The predicted values for the lifetime cancer risk of a given tissue type are given by the x-axis coordinate for that tissue, while the actual observed values in the U.S. for those cancer types are provided by the y-axis coordinate. The red line represents a perfect fit between the predicted and observed values (slope =1). The predicted values are based on conservatively assumptions of the Armitage and Doll model, i.e. where clonal expansions are not considered, and on the assumption that only replicative mutations are present (no environmental and inherited factors). The inclusion of clonal expansions in the mathematical model, or the inclusion of extrinsic factors, would shifts points further to the right. As depicted in the figure, for many tissues, random replicative mutations are sufficient to explain the observed incidences.

Driver gene mutation probability per cell division = $5 \times 10^{-7}$. The number of required drivers was assumed to be 3 except when available evidence suggested that only two driver gene mutations were required (osteosarcomas (pelvis, arms, head, legs, overall), ovarian germ cell, glioblastoma, medulloblastoma, gallbladder, and medullary and follicular thyroid cancers). In the Armitage and Doll formula, time was substituted with the estimated number of cell turnovers.

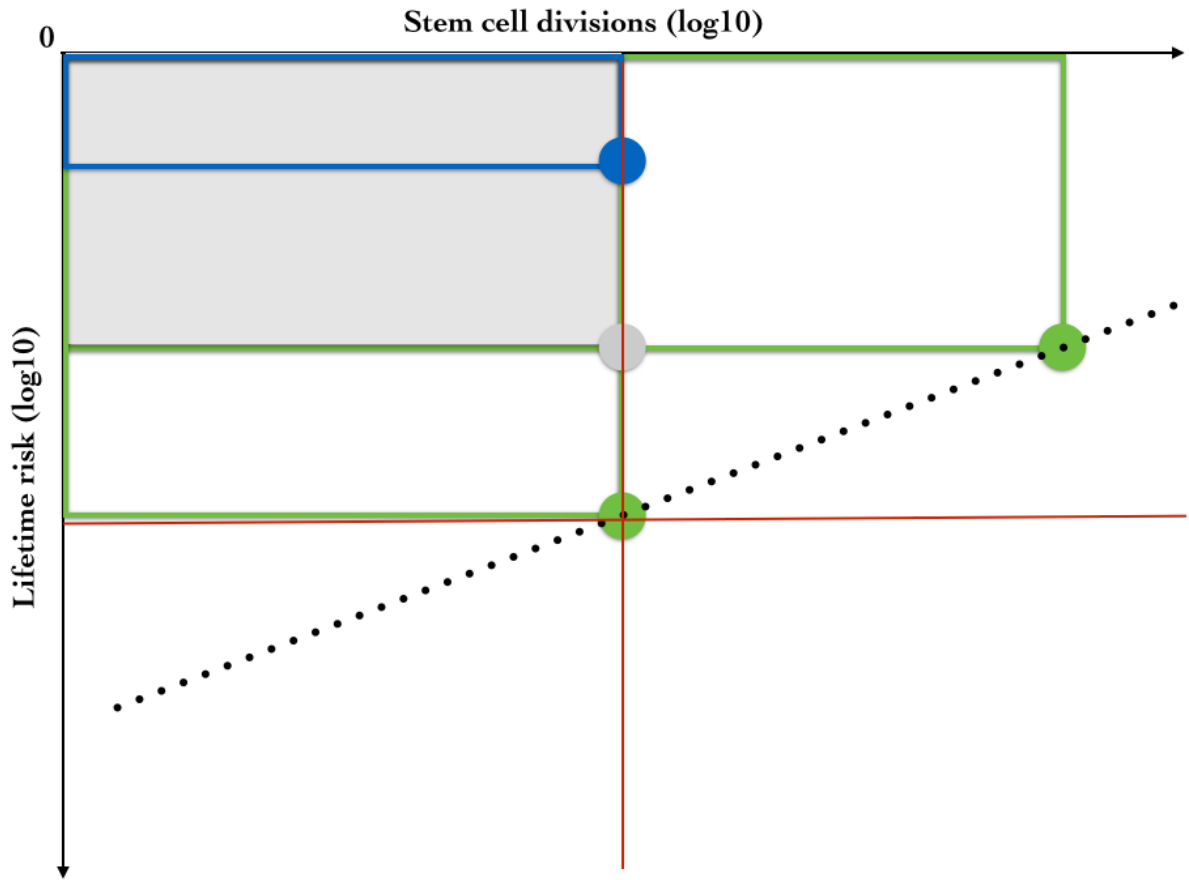

**Figure TR3. An illustration of the idea behind the Extra Risk Score.** See Point 6 in the text for a detailed explanation.

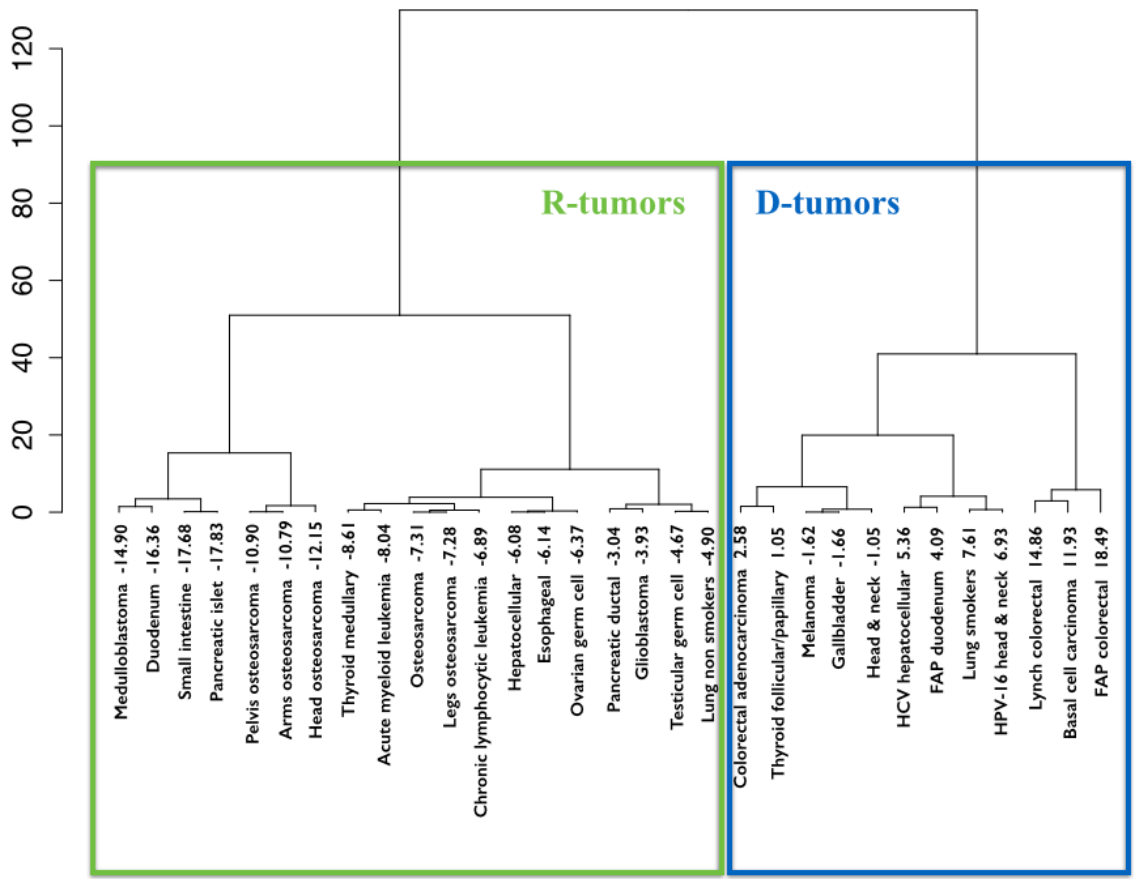

**Figure S1. Ward hierarchical clustering of D- and R-tumors.** The difference with the K-means clustering presented in Figure 2 of (Tomasetti and Vogelstein, 2015), is that now three R-tumors closest to the D-tumors shifted to the D- tumor cluster.